# River environmental restoration based on random observations of a non-smooth stochastic dynamical system

Hidekazu Yoshioka[1, 2, *]

[1] Assistant Professor, Graduate School of Natural Science and Technology, Shimane University, Nishikawatsu-cho 1060, Matsue, 690-8504, Japan

[2] Center Member, Fisheries Ecosystem Project Center, Shimane University, Nishikawatsu-cho 1060, Matsue, 690-8504, Japan

* Corresponding author
E-mail: yoshih@life.shimane-u.ac.jp

**Abstract**

Earth and soils are indispensable elements of river environment. Dam-downstream environment and ecosystems have been severely affected by reduced or even stopped sediment supply from the upstream. Replenishing earth and soils from outside the river has been considered as an effective way to mitigate this issue. However, its cost-effective implementation has not been considered from a theoretical side. This paper presents a tractable new stochastic control model to deal with this issue. The sediment dynamics in the river environment follow non-smooth and continuous-time piecewise deterministic dynamics. The model assumes that the observation of the sediment dynamics is carried out only randomly and discretely, and that the sediment can be replenished at each observation time with cost. This partial observation assumption is consistent with the fact that continuously obtaining the environmental information is difficult in applications. The performance index to penalize the sediment depletion has a non-smooth term as well. We demonstrate that these non-smoothness factors harmonize with a dynamic programming principle, and derive the optimality equation in a degenerate elliptic form governing the most cost-efficient sediment replenishment policy. We analytically derive and verify an exact solution under a simplified condition for a discounted case, an Ergodic case, and a complete information case. A more realistic case is handled using a high-resolution finite difference scheme. We then provide the optimal sediment replenishment policy numerically.

## 1. Introduction

### 1.1 Problem background

Harmonization between river environment and human activities is a long-standing issue (Briones-Hidrovo et al., 2020; Murphy, 2019; Han et al., 2020). River environment and ecology in dam-downstream reaches have severely been affected by operating dams, although they have been playing central roles in multiple industrial purposes like water resources management and hydropower production, and disaster prevention (Briones-Hidrovo et al., 2019; de Assis Espécie et al., 2019). The regulated flow conditions resulting from dam operations usually have less flow fluctuations compared with the natural ones (Mori et al., 2018), and negatively affect sediment dynamics (Nukazawa et al., 2020), water quality dynamics (Rolls et al., 2020), ecological dynamics (Baumgartner et al., 2020), and biodiversity (Xu et al., 2020).

Among many environmental problems, a common issue that dam-downstream river environments worldwide encounter is the sediment trapping by dams (Walling and Fang, 2003; Zhang et al., 2019), with which the sediment transport from the upstream toward downstream is critically reduced. As a mitigation policy against the sediment trapping, replenishment of earth and soils from outside the river has been carried out in several case studies (Schleiss et al., 2016; Brousse et al., 2019; Stähly et al., 2019). It has been experimentally found that this is an effective strategy to restore the dam-downstream river environment. In fact, sediment particles contribute to flushing out of the nuisance benthic filamentous algae found in dam-downstream rivers (Fovet et al., 2010; Luce et al., 2010; Katz et al., 2018; Neverman et al., 2018).

However, optimization of the sediment replenishment from a theoretical cost-effectiveness viewpoint has been paid less attention. Especially, mathematical modeling for cost-efficient sediment replenishment policies have not been discussed to the best of the authorøs knowledge except for several recent research (Yoshioka et al., 2019a-b). Mathematically analyzing the sediment dynamics in the dam-downstream rivers along with sediment replenishment policy would deepen understanding of the problem. In addition, such a mathematical model would be able to provide useful insights into real problems.

### 1.2 Mathematical background

River flow regimes are reasonably represented as continuous-time stochastic processes where physically unresolved components are efficiently described with Markovian noises (Song et al., 2020; Tu et al., 2019; Ramirez and Constantinescu, 2020). In this view, sediment storage dynamics in river environment can be described as some stochastic differential equations (SDEs) (Øksendal and Sulem, 2019) driven by stochastic processes representing river flow regimes. Then, the optimal control of sediment replenishment from the outside to the river environment is formulated as a stochastic control problem of a system of SDEs (Øksendal and Sulem, 2019). The previous research results suggest that the sediment replenishment is described as an impulsive intervention (Yoshioka et al., 2019a-b). The target problem can therefore be formulated as a stochastic impulse control problem.

There exist two potential difficulties in this control problem. The first one is the system non-smoothness. Physically, the sediment transport takes place when the bottom shear stress, which is an increasing function of the river discharge, exceeds a threshold value (Ancey, 2020). In addition, the transport rate, which is the amount of sediment transported toward the downstream per unit time, is an increasing function of the flow discharge. The transport rate of a sediment lump placed in a flowing river is then positive assuming that the flow discharge is sufficiently large. We then encounter the dynamics

$$\mathrm{d}Y_t = -S\chi_{\{Y_t>0\}}\mathrm{d}t \quad \text{for} \quad t>0, \quad Y_0>0 \tag{1}$$

where $t$ is time, $Y = (Y_t)_{t\geq 0}$ is the volume of the sediment lump stored in a river reach, $S>0$ is the transport rate per unit time, and $\chi_A$ is the indicator function for the set $A$.

The model parameters in (1) are assumed to be time-independent for convenience, but similar dynamics emerge under a stochastic environment that we consider later. A potential issue is that the dynamics are non-smooth due to the appearance of $\chi_{\{Y_t>0\}}$. Owing to the one-sided Lipschitz continuity of the right-hand side of (1), the unique Filippov solution is (Cortes, 2008):

$$Y_t = \max\{Y_0 - St, 0\}, \tag{2}$$

which is not differentiable at the time $t = S^{-1}Y_0$. A similar non-smoothness is encountered under a stochastic environment with a random $S$ as well. Therefore, we must consider an optimal control problem of non-smooth dynamics. Non-smooth dynamics have richer mathematical structures than the smooth one (Kim and Wang, 2018; Noori Skandari and Ghaznavi, 2017; Chen et al., 2019; Mertz et al., 2019; Mertz and Pironneau, 2019), and are therefore potentially more complicated. As we will see, the non-smoothness is inherited in control problems of the dynamics.

The second difficulty is the incompleteness of information. It is not always possible to obtain environmental information continuously in time, but only discretely in real problems (Yoshioka and Tsujimura, 2020). Therefore, we must deal with a stochastic control problem under partial (discretely sampled) information. In addition, completely scheduled observations may not be easy because an environmental manager has multiple tasks in general. We therefore assume discrete and random observations (Wang, 2001; Pham and Tankov, 2008) where the information available for making decisions is smaller than that under the complete information setting (Korn et al., 2017; Wu, 2019). Stochastic control under discrete random observations have been studied theoretically under finite-horizon (Pham and Tankov, 2008), discounted infinite-horizon (Wang, 2001), and Ergodic (Menaldi and Robin, 2017) settings. Federico and Gassiat (2014) considered an extended problem controlling SDEs governing both completely observable and only discretely observable variables. Regime-switching problems have also been considered (Capponi and Figueroa-López, 2014; Wei et al., 2012). The random observation framework also covers optimal stopping problems (Boyarchenko and Levendorski , 2019). As reviewed above, the stochastic control under discrete and random observations have been studied so far; however, most of them are related to problems of finance and economics, and far less approached problems related to environment and ecology. In addition, they handle only smooth dynamics.

In summary, the stochastic control problem of the sediment dynamics in a dam-downstream river environment can be formulated as a stochastic impulse control problem of a non-smooth system under discrete and random observations. This kind of modeling has not been studied so far except for our recent research (Yoshioka et al., 2020a), despite it is related to the important engineering problems arising in environmental management. The previous model (Yoshioka et al., 2020a) was considering coupled biological (algae growth) and environmental dynamics (flow regimes and sediment storage), and was too complicated to analyze theoretically. Mathematically modeling and theoretically analyzing this engineering problem would contribute to deepening connections between stochastic control and environmental management and further provide useful insights into real problems.

**1.3 Our contribution**

We demonstrate that finding a cost-efficient sediment replenishment policy is viewed as a stochastic control problem under partial observations. An SDE describing the lumped sediment storage dynamics in a dam-downstream river is formulated as a more realistic counterpart of (1). Its well-posedness is then discussed, and a stochastic impulse control problem based on discrete and random observations with sediment replenishment is formulated. Other than the system dynamics, the performance index to be minimized by a cost-efficient sediment replenishment policy is also non-smooth to penalize the sediment depletion. This is also a unique point in our model. We show that these non-smoothness factors harmonize with the physically-based non-smoothness of the system dynamics.

We derive and verify an exact solution under a simplified condition for a discounted case, an Ergodic case, and a complete information case. We can then more clearly understand the parameter dependence of the optimal sediment replenishment policy. The full problem without the above simplification is analyzed from a viewpoint of smooth solutions as well. We therefore show the existence of a reasonable smooth solution to the non-smooth degenerate elliptic system. Several related issues like model uncertainty, delayed execution, and approaches from a viscosity viewpoint are discussed as well. The model is finally applied to finding the cost-efficient sediment replenishment policy under stochastic flow regimes identified at a river. The computational results validate key assumptions made in the mathematical analysis. Our contributions are therefore formulation, mathematical analysis, computation, and application of the new stochastic control model with partial observation.

### 1.4 Organization of this paper

The rest of this manuscript is organized as follows. We present the dynamics and its control problem in Section 2. An exactly-solvable problem and the full problem are analyzed in Section 3. Several related issues are discussed as well in this section. The full problem is numerically computed in Section 4 using identified parameter values. Summary and future perspectives of this paper are presented in Section 5.

## 2. Mathematical model

### 2.1 Un-controlled dynamics

The system dynamics in our model contain a continuous-time and right-continuous Markov chain with the left-limit representing the river flow regimes in a dam-downstream river reach and the lumped sediment storage dynamics in the reach. We employ a standard complete probability space as in the conventional stochastic control framework (Øksendal and Sulem, 2019).

The time is denoted as $t \geq 0$. The flow discharge is assumed to follow a continuous-time finite-regime Markov chain $\alpha = (\alpha_t)_{t \geq 0}$ (Yin and Zhu, 2009; Yoshioka et al., 2020a). The total number of the regimes is denoted as $I \in \mathbb{N}$ and the associated constant switching matrix as $\nu = [\nu_{i,j}]_{1 \leq i,j \leq J}$ whose entries are bounded and non-negative. Set $M = \{i\}_{1 \leq i \leq I}$. The river discharge is denoted as $q_i > 0$ for the $i$ th regime. Without any loss of generality, the sequence $\{q_i\}_{1 \leq i \leq I}$ is set to be strictly increasing: $q_i < q_j$ for $i < j$ ($i, j \in M$). The Markov chain $\alpha$ is assumed to admit a unique stationary distribution.

The transport rate, which is the amount of sediment transported toward the downstream per unit time, depends on the flow regime. Its functional form will be specified in the application presented later. What is important at the theoretical level is that the transport rate vanishes if the discharge is smaller than a threshold. This threshold is determined by physical factors such as the river geometry and sediment particle diameter (Ancey, 2020). In this view, the transport rate is a bounded function $S: \mathbb{R}_+ \to \mathbb{R}_+$ with $S(q_I) > 0$. We assume

$$\int_0^t S(q_{\alpha_{s-}}) \mathrm{d}s \to +\infty \quad \text{almost surely (a.s.), when } t \to +\infty, \tag{3}$$

so that the sediment depletion ($Y_t = 0$) occurs at some $t \geq 0$. Otherwise, replenishing the sediment may be unnecessary and the control problem in this paper becomes trivial.

The SDE governing the sediment storage dynamics is set as

$$\mathrm{d}Y_t = -S(q_{\alpha_{t-}}) \chi_{\{Y_{t-} > 0\}} \mathrm{d}t \quad \text{for } t > 0, \quad Y_0 \geq 0. \tag{4}$$

We assume that the maximum amount of sediment allowed to be stored in the reach equals 1 by a normalization. Therefore, the range of the stored sediment is $D = [0,1]$. We thus require $Y_0 \in D$. A candidate of continuous solutions to the SDE (4) is expressed as

$$Y_t = \max\left\{0, Y_0 - \int_0^t S(q_{\alpha_{s-}}) \mathrm{d}s\right\} \quad \text{for } t \geq 0, \tag{5}$$

which is a.s. continuous with respect to $t$ and non-negative. This solution is non-increasing for $t \geq 0$ because $S \geq 0$.

We see that the SDE(4) has a unique path-wise solution expressed as (5) despite it has a non-Lipschitz continuous coefficient.

*Proposition 1*

*There exists a unique continuous path-wise solution to the SDE(4) a.s. The solution is expressed as (5).*

**(Proof of Proposition 1)**

The existence is straightforward. The uniqueness is proven as follows. Assume that there exist two solutions $Y^{(1)}, Y^{(2)}$ satisfying (4). Applying Itô's formula to $Z = \frac{1}{2}(Y^{(1)} - Y^{(2)})^2$ yields

$$\mathrm{d}Z_t = (Y_t^{(1)} - Y_t^{(2)}) \mathrm{d}(Y_t^{(1)} - Y_t^{(2)}) = -S(q_{\alpha_{t-}})(Y_t^{(1)} - Y_t^{(2)})(\chi_{\{Y_t^{(1)} > 0\}} - \chi_{\{Y_t^{(2)} > 0\}}) \mathrm{d}t. \tag{6}$$

We get

$$Z_t = -\int_0^t S(q_{\alpha_{s-}})(Y_s^{(1)} - Y_s^{(2)})(\chi_{\{Y_s^{(1)} > 0\}} - \chi_{\{Y_s^{(2)} > 0\}}) \mathrm{d}s, \quad t \geq 0. \tag{7}$$

For $y_1, y_2 \in \mathbb{R}$, we see

$$(y_1 - y_2)(\chi_{\{y_1 > 0\}} - \chi_{\{y_2 > 0\}}) \geq 0. \tag{8}$$

By (7) and (8), we get

$$Z_t = \frac{1}{2}\left(Y_t^{(1)} - Y_t^{(2)}\right)^2 \leq 0, \quad t \geq 0, \tag{9}$$

with which we see the uniqueness $Y_t^{(1)} = Y_t^{(2)}$ a.s. $t \geq 0$.

*Remark 1*

One may alternatively formulate smooth dynamics using some regularization (Serdukova et al., 2017; Yoshioka et al., 2019b). We do not use the regularization so that the depleted state ($Y_t = 0$) is not dispersed.

*Remark 2*

The algae population dynamics (Yoshioka, 2019) can be coupled with the sediment storage dynamics. Logistic models (Brites and Braumann, 2020) may be such candidates. Fisheries dynamics (Yoshioka et al., 2019b) can also be coupled with the presented models. However, we solely focus on the sediment dynamics to formulate a simpler model.

**2.2 Controlled dynamics**

We assume that the decision-maker, the environmental manager, controls the sediment storage dynamics based on discrete observations. The observation process is assumed to be random, and is expressed as a standard Poisson process $N = (N_t)_{t \geq 0}$ with the intensity $\lambda > 0$. The sequence of observation times is expressed as an increasing sequence $\tau = \{\tau_k\}_{k \in \mathbb{N}}$, which collects the jump times of $N$. Without any loss of generality, set $\tau_0 = 0$. The Poisson nature of the observation times is assumed to make the model simpler; however, we are not sure that the model without the Poisson nature of the observation process can be handled within the framework presented in this paper. Notice that the switching times of the Markov chain $\alpha$ and the observation times do not coincide with each other a.s., because of their Poisson nature.

We assume that the decision-maker can know the realization $\left(\alpha_{\tau_k}, Y_{\tau_k}\right)$ at each $\tau_k$. A natural filtration generated by the available information is then set as $\mathcal{F} = (\mathcal{F}_t)_{t \geq 0}$, where

$$\mathcal{F}_t = \sigma\left\{\left(\tau_j, \alpha_{\tau_j}, Y_{\tau_j}\right)_{0 \leq j \leq k}, k = \sup\{j : \tau_j \leq t\}\right\}, \quad t \geq 0 \tag{10}$$

because only the discretely sampled information is available for the decision-maker.

We assume that the decision-maker can supply earth and soils from outside the river at each observation. We consider the following policy at each $\tau_i$:

$$Y_{\tau_k+} = Y_{\tau_k} + \eta_k \text{ with } \eta_k = \begin{cases} 0 & (\text{Do nothing}) \\ 1 - Y_{\tau_k} & (\text{Replenish}) \end{cases}, \quad (11)$$

where $\eta_k$ represents the amount of sediment supplied at $\tau_k$. Therefore, we are assuming that the decision-maker does nothing or supplies the sediment to the maximum level 1 at each observation. Without any loss of generality, set $\eta_0 = 0$.

The controlled sediment storage dynamics are described as

$$dY_t = -S(q_{\alpha_{t-}})\chi_{\{Y_{t-}>0\}}dt + \bar{\eta}_t dN_t \text{ for } t > 0, \; Y_0 \geq 0, \quad (12)$$

where the process $\bar{\eta} = (\bar{\eta}_t)_{t \geq 0}$ equals $\eta_k$ at $\tau_k$ and equals 0 otherwise. By (11) and **Proposition 1**, we see that the process $Y$ is a.s. confined in $D$. We understand the product term $\bar{\eta}_t dN_t$ at $\tau_k$ as $\eta_{\tau_k}$. A set of admissible control $C$ contains a continuous-time processes $\bar{\eta} = (\bar{\eta}_t)_{t \geq 0}$ such that it equals $\eta_k$ at $\tau_k$ and 0 otherwise, where $\eta_k$ is measurable with respect to $\mathcal{F}_{\tau_k}$ with (11).

## 2.3 Performance index and value function

A performance index as a metric to be minimized with respect to $\bar{\eta} \in C$ is presented. Since the sediment depletion ($Y_t = 0$) critically affects the river environment and ecology, its occurrence should be penalized. To avoid the occurrence of the depletion, the decision-maker can supply sediment from outside the river; however, such an activity can be costly. The cost of sediment would be incurred per unit volume or weight (proportional cost), and some labor costs to transport the sediment would be necessary as well (fixed cost).

For convenience, set $\Omega = M \times D$. The conditional expectation with respect to $(\alpha_0, Y_0) = (i, y)$ is denoted as $\mathbb{E}^{i,y}$. Considering the penalization and replenishment costs, we set the performance index $\phi: \Omega \times C \to \mathbb{R}$ as

$$\phi(i, y, \bar{\eta}) = \mathbb{E}^{i,y}\left[\int_0^\infty e^{-\delta s}\chi_{\{Y_{s-}=0\}}ds + \sum_{k \geq 1} e^{-\delta \tau_k}\left(c\eta_k + d\chi_{\{\eta_k > 0\}}\right)\right], \quad (13)$$

where $\delta > 0$ is the discount rate, $c > 0$ is the coefficient of proportional cost, and $d > 0$ is the fixed cost. This performance index represents the discounted sum of the penalization term against the sediment depletion and the costs of sediment replenishment. The sediment dynamics are penalized when the sediment storage depletes ($Y_t = 0$). By (1) and the Markov property of the river flow regime, we infer that the occurrence probability of such an event is not null. Later, we show that introducing the discount rate allows us to handle both the discounted infinite-horizon problem and an Ergodic limit; the latter is a vanishing discount limit ($\delta \to 0$).

The value function $\Phi : \Omega \to \mathbb{R}$ is the minimized $\phi$ with respect to $\bar{\eta} \in C$:

$$\Phi(i,y) = \inf_{\bar{\eta} \in C} \phi(i,y,\bar{\eta}). \tag{14}$$

This is a non-negative function by the non-negativity of each term in the right-hand side of (13). It is bounded in $\Omega$. In fact, choosing the null-control $\bar{\eta}_0$ with $\eta_i = 0$ ($i \geq 0$) yields

$$0 \leq \Phi(i,y) \leq \phi(i,y,\bar{\eta}_0) = \mathbb{E}^{i,y}\left[\int_0^\infty e^{-\delta s}\chi_{\{Y_{s-}=0\}}\mathrm{d}s\right]_{\bar{\eta}=\bar{\eta}_0} \leq \frac{1}{\delta} < +\infty, \quad (i,y) \in \Omega. \tag{15}$$

Therefore, the value function is well-defined.

An optimal policy $\eta^* = \left(\eta_t^*\right)_{t \geq 0}$ as an element of $C$ minimizing $\phi$ is referred to as the optimal control. The goal of our control problem is to find $\eta^*$ based on the observables.

## 2.4 Optimality equation

The optimality equation is the equation governing the value function $\Phi$. This equation is formally derived by applying a dynamic programming principle to $\Phi$, which is valid if $\Phi$ is continuously differentiable with respect to the second argument in $\Omega$. Such an assumption is not always satisfied because the optimality equation is of a degenerate elliptic form as shown below. A degenerate elliptic equation does not always have a classical solution satisfying the equation pointwise sense, and we must often use a weaker notion of solutions like viscosity solutions (Crandall et al., 1992). Later, we show that its reasonable solution is smooth.

For our model, the optimality equation is formally derived as (Wang, 2001)

$$\begin{aligned}&\delta\Phi_i + S(q_i)\chi_{\{y>0\}}\frac{\mathrm{d}\Phi_i}{\mathrm{d}y} + \sum_{i \neq j \in M} v_{i,j}\left(\Phi_i - \Phi_j\right) \\ &+ \lambda\left(\Phi_i - \inf_{\eta \in C(y)}\left\{\Phi_i(y+\eta) + c\eta + d\chi_{\{\eta>0\}}\right\}\right) - \chi_{\{y=0\}} = 0\end{aligned}, \quad (i,y) \in \Omega, \tag{16}$$

where we used the notation $\Phi_i(y) = \Phi(i,y)$ and $C(\cdot)$ is the state-dependent binary set

$$C(y) = \{0, 1-y\}, \quad y \in D. \tag{17}$$

The optimality equation (16) is considered on $\Omega$ without specifying boundary conditions. On the boundary $y = 0$, (16) reduces to

$$\delta\Phi_i + \sum_{i \neq j \in M} v_{i,j}\left(\Phi_i - \Phi_j\right) + \lambda\left(\Phi_i - \inf_{\eta \in C(0)}\left\{\Phi_i(\eta) + c\eta + d\chi_{\{\eta>0\}}\right\}\right) - 1 = 0, \tag{18}$$

which is interpreted as a non-local boundary condition connecting the information between $y = 0, 1$. This non-locality comes from the assumed sediment replenishment policy.

Our optimality equation (16) is a system of nonlinear and nonlocal degenerate elliptic equations. Its justification is discussed in the next section, where we show that the optimal sediment replenishment policy is constructed using $\Phi$. In this sense, solving the optimal control problem is reduced to finding an appropriate solution to (16).

Formally, assuming a Markov control of the form $\eta^* = \eta^*\left(\alpha_{\tau_k}, Y_{\tau_k}\right)$ at each $\tau_k$, we can guess the functional form of the optimal control using the observables $\left(\alpha_{\tau_k}, Y_{\tau_k}\right)$ as

$$\eta^*\left(\alpha_{\tau_k}, Y_{\tau_k}\right) = \arg\min_{\eta \in C(Y_{\tau_k})} \left\{ \Phi\left(\alpha_{\tau_k}, Y_{\tau_k} + \eta\right) + c\eta + d\chi_{\{\eta>0\}} \right\}. \qquad (19)$$

This minimizer exists because $C$ in (17) is compact. From a technical point, it is useful to see the technical relationship

$$\Phi_i(y) - \inf_{\eta \in C(y)} \left\{ \Phi_i(y+\eta) + c\eta + d\chi_{\{\eta>0\}} \right\} = 0 \quad \text{if} \quad \eta^*(i,y) = 0, \quad (i,y) \in \Omega, \qquad (20)$$

meaning that we may simplify the optimality equation when $\eta^*(i,y) = 0$. This relationship will be utilized in the next section.

## 3. Mathematical analysis

### 3.1 An exactly-solvable case

Here, we analyze an exactly-solvable case of the stochastic control problem under a simplified condition. The exact solution is non-trivial and provides much information on the controlled dynamics. We can more clearly understand parameter dependence of the optimal control by analyzing the exactly-solvable case. Furthermore, the solution can be utilized as a benchmark to verify accuracy of numerical schemes. Due to the simplicity, we can verify the optimality of the solution directly. We thus demonstrate the existence of a reasonable smooth solution to a non-smooth degenerate elliptic system.

#### 3.1.1 Discounted case

The problem here considers a single-regime case ($I = 1$). The subscripts representing the regimes are omitted in this sub-section for convenience. We assume that the transport rate is a constant $S > 0$. The optimality equation governing the value function $\Phi = \Phi(y)$ in this case becomes

$$\delta\Phi + S\chi_{\{y>0\}} \frac{d\Phi}{dy} + \lambda\left( \Phi - \inf_{\eta \in C(y)} \left\{ \Phi(y+\eta) + c\eta + d\chi_{\{\eta>0\}} \right\} \right) - \chi_{\{y=0\}} = 0, \quad y \in D. \qquad (21)$$

This is still a non-linear and non-local degenerate elliptic equation. We explore a solution corresponding to the following threshold type control

$$\eta^*_{\tau_k} = \begin{cases} 1-Y_{\tau_k} & (0 \leq Y_{\tau_k} \leq \bar{y}) \\ 0 & (\bar{y} < Y_{\tau_k} \leq 1) \end{cases} \quad (22)$$

with a threshold value $\bar{y} \in (0,1)$. This policy means that the sediment supply should be carried out if the sediment storage is smaller than $\bar{y}$. This policy is reasonable because it is natural to seek for a policy to increase the storage when it is small. The threshold value $\bar{y}$ depends on the parameter values like river environmental conditions and the incurred cost and penalization.

We construct a candidate solution $\Psi \in C^1(D)$ and verify that the solution is the value function $\Phi$ such that the policy (22) is optimal. This solution, if it exists, is a classical solution satisfying the optimality equation pointwise. By (20), we rewrite (21) as

$$\delta\Psi + S\chi_{\{y>0\}}\frac{d\Psi}{dy} + \lambda\left(\Psi - \inf_{\eta \in C(y)}\{\Psi(y+\eta) + c\eta + d\chi_{\{\eta>0\}}\}\right) - \chi_{\{y=0\}} = 0, \quad 0 \leq y \leq \bar{y} \quad (23)$$

and

$$\delta\Psi + S\frac{d\Psi}{dy} = 0, \quad \bar{y} < y \leq 1. \quad (24)$$

From (23)-(24), we get

$$\Psi(y) = \begin{cases} fe^{-\frac{\delta+\lambda}{S}y} + ay + b & (0 \leq y \leq \bar{y}) \\ \Psi(1)e^{\frac{\delta}{S}(1-y)} & (\bar{y} < y \leq 1) \end{cases} \quad (25)$$

with constants

$$a = -\frac{\lambda c}{\delta+\lambda}, \quad b = \frac{1}{\delta+\lambda}(-aS + \lambda(\Psi(1)+c+d)), \text{ and } f = \frac{\delta+\lambda-\lambda cS}{(\delta+\lambda)^2}. \quad (26)$$

Notice the dependence of $b$ on $\Psi(1)$. The fact that $\Psi$ in (25) satisfies (23)-(24) can be checked directly.

There are the two unknowns $\bar{y}$ and $\Psi(1)$. Assume $\bar{y} \in (0,1)$. We determine them by requiring smoothness of the solution at $y = \bar{y}$:

$$\Psi(\bar{y}-0) = \Psi(\bar{y}+0) \text{ and } \frac{d\Psi}{dy}(\bar{y}-0) = \frac{d\Psi}{dy}(\bar{y}+0) \quad (27)$$

or equivalently

$$fe^{-\frac{\delta+\lambda}{S}\bar{y}} + a\bar{y} + b = \Psi(1)e^{\frac{\delta}{S}(1-\bar{y})} \text{ and } -\frac{\delta+\lambda}{S}fe^{-\frac{\delta+\lambda}{S}\bar{y}} + a = -\frac{\delta}{S}\Psi(1)e^{\frac{\delta}{S}(1-\bar{y})} \quad (28)$$

with (26). We have the two equations for determining the two unknowns.

The next proposition shows that the guessed smooth solution $\Psi$, if it exists, is the value function $\Phi$ and the control (22) is optimal. Its existence is discussed in **Remark 3**.

*Proposition 2*

*Assume that the system (28) admits a solution $(\Psi(1), \bar{y})$ such that $0 < \bar{y} < 1$. Then, we get the optimally-controlled dynamics*

$$dY_t = -S\chi_{\{Y_{t-}>0\}}dt + (1-Y_t)\chi_{\{Y_{t-}\leq \bar{y}\}}dN_t, \qquad (29)$$

*which gives the value function $\Phi$. Namely, the policy (22) is optimal.*

**(Proof of Proposition 2)**

We can follow the proof of Theorem 1 in Wang (2001) with several modifications because the underlying dynamics and performance indices are different. Set some $y \in D$. By Itô's formula for $\psi \in C^1(D)$ and a policy $\bar{\eta} \in C$, we get

$$\begin{aligned}e^{-\delta T}\psi(Y_T) - \psi(y) &= \int_0^T e^{-\delta s}\left(-\delta\psi(Y_{s-}) - S\chi_{\{Y_{s-}>0\}}\psi'(Y_{s-})\right)ds \\ &\quad + \sum_{0\leq t\leq T}e^{-\delta t}\left(\psi(Y_t + \bar{\eta}_t dN_t) - \psi(Y_t)\right)\end{aligned}, \quad T>0. \qquad (30)$$

By $\Psi \in C^1(D)$ and (21), we have

$$\begin{aligned}e^{-\delta T}\Psi(Y_T) - \Psi(y) &= \int_0^T e^{-\delta s}\chi_{\{0<Y_{s-}\leq 1\}}\left(-\delta\Psi(Y_{s-}) - S\chi_{\{Y_{s-}>0\}}\Psi'(Y_{s-})\right)ds \\ &\quad + \int_0^T e^{-\delta s}\chi_{\{Y_{s-}=0\}}\left(-\delta\Psi(Y_{s-}) - S\chi_{\{Y_{s-}>0\}}\Psi'(Y_{s-})\right)ds \\ &\quad + \sum_{0\leq t\leq T}e^{-\delta t}\left(\Psi(Y_t + \bar{\eta}_t dN_t) - \Psi(Y_t)\right) \\ &= \int_0^T e^{-\delta s}\lambda\left(\Psi(Y_{s-}) - \inf_\eta\left\{\Psi(Y_{s-}+\eta) + c\eta + d\chi_{\{\eta>0\}}\right\}\right)ds \\ &\quad + \int_0^T e^{-\delta s}\chi_{\{Y_{s-}=0\}}(-1)ds + \sum_{0\leq t\leq T}e^{-\delta t}\left(\Psi(Y_t + \bar{\eta}_t dN_t) - \Psi(Y_t)\right)\end{aligned} \qquad (31)$$

with the notation $\Psi' = \dfrac{d\Psi}{dy}$. Since $\bar{\eta}_t \neq 0$ only at jumps, we have

$$\Psi(Y_t + \bar{\eta}_t dN_t) - \Psi(Y_t) \geq g(Y_t)dN_t - \left(c + d\chi_{\{\bar{\eta}_t>0\}}\right) \qquad (32)$$

with $g: D \to \mathbb{R}$ given by

$$g(y) = \inf_\eta\left\{\Psi(y+\eta) + c\eta + d\chi_{\{\eta>0\}}\right\} - \Psi(y). \qquad (33)$$

Substituting (32) into (31) with the help of (33) yields

$$e^{-\delta T}\Psi(Y_T)-\Psi(y)=-\int_0^T e^{-\delta s}\chi_{\{Y_{s-}=0\}}\,\mathrm{d}s-\int_0^T e^{-\delta s}\lambda g(Y_{s-})\,\mathrm{d}s$$
$$+\sum_{0\le t\le T}e^{-\delta t}\left(\Psi(Y_t+\bar{\eta}_t\mathrm{d}N_t)-\Psi(Y_t)\right)$$
$$\ge -\int_0^T e^{-\delta s}\chi_{\{Y_{s-}=0\}}\,\mathrm{d}s-\int_0^T e^{-\delta s}\lambda g(Y_{s-})\,\mathrm{d}s \qquad (34)$$
$$+\sum_{0\le t\le T}e^{-\delta t}\left\{g(Y_t)\mathrm{d}N_t-\left(c+d\chi_{\{\eta_t>0\}}\right)\right\}$$

Notice that the compensated Poisson process $\tilde{N}_t=N_t-\lambda t$ $(t\ge 0)$ is a Martingale. Taking the expectation $\mathbb{E}^y$ in both sides of (34) and rearranging it then leads to

$$\Psi(y)\le e^{-\delta T}\mathbb{E}^y\left[\Psi(Y_T)\right]+\mathbb{E}^y\left[\int_0^T e^{-\delta s}\chi_{\{Y_{s-}=0\}}\,\mathrm{d}s+\sum_{0\le t\le T}e^{-\delta t}\left(c+d\chi_{\{\eta_t>0\}}\right)\right]$$
$$-\mathbb{E}^y\left[\int_0^T e^{-\delta s}g(Y_s)\,\mathrm{d}\tilde{N}_s\right] \qquad (35)$$

By the Martingale property of $\tilde{N}$ and the smoothness of $\Psi$, we get

$$\mathbb{E}^y\left[\int_0^T e^{-\delta s}g(Y_s)\,\mathrm{d}\tilde{N}_s\right]=0. \qquad (36)$$

Taking the limit $T\to+\infty$ in (35) with the help of (36) yields

$$\Psi(y)\le \mathbb{E}^y\left[\int_0^\infty e^{-\delta s}\chi_{\{Y_{s-}=0\}}\,\mathrm{d}s+\sum_{k\ge 1}e^{-\delta\tau_k}\left(c\eta_k+d\chi_{\{\eta_k>0\}}\right)\right] \qquad (37)$$

since $\Psi$ is uniformly bounded in $D$. Because $\bar{\eta}\in\mathcal{C}$ is arbitrary, we get

$$\Psi\le\Phi \text{ in } D. \qquad (38)$$

Next, we show the equality

$$\Psi=\Phi \text{ in } D, \qquad (39)$$

with which we can complete the proof. We only need to prove $\Psi\ge\Phi$. This immediately follows from the fact that the control (22) is admissible. Especially, (35) becomes the equality

$$\Psi(y)=e^{-\delta T}\Psi(Y_T)+\left(\int_0^T e^{-\delta s}\chi_{\{Y_{s-}=0\}}\,\mathrm{d}s+\sum_{0\le t\le T}e^{-\delta t}\left(c+d\chi_{\{\eta_t>0\}}\right)\right)-\int_0^T e^{-\delta s}g(Y_s)\,\mathrm{d}\tilde{N}_s \qquad (40)$$

with this threshold type control. We again take the expectation $\mathbb{E}^y$ and then the limit $T\to+\infty$. The proof is completed.

*Remark 3*

A central assumption in **Proposition 2** is whether the system (28) admits a solution $(\Psi(1),\bar{y})$ such that $0<\bar{y}<1$. The Ergodic case analyzed below suggests the existence of $0<\bar{y}<1$ for small $c,d,\delta>0$. This is because the terms in (28) depend smoothly on $c,d,\delta>0$.

### 3.1.2 Ergodic case

The exactly-solvable case can be further reduced assuming an Ergodic limit, which is a long-time limit without any discounting. As demonstrated here, the coefficients of the value function are determined analytically in this case. The system (28) reduces to a simpler couple of equations whose unique solvability is established under certain conditions.

We consider the Ergodic limit $\delta \to +0$ where we formally assume that the value function $\Phi$ multiplied by $\delta$ converges toward the effective Hamiltonian $u$, which formally corresponds to the small-$\delta$ limit (Qian, 2003):

$$u \to \delta\Phi \quad \text{in} \quad D \quad \text{as} \quad \delta \to +0. \tag{41}$$

Firstly, we again assume $\bar{y} \in (0,1)$. This assumption is justified later. Taking this the limit $\delta \to +0$ in the second equation of (28) yields

$$(1-cS)e^{-\frac{\lambda}{S}\bar{y}} + cS = u. \tag{42}$$

Taking the limit in the first equation of (28) is a bit more complicated. Rewrite it with (26) as

$$fe^{-\frac{\delta+\lambda}{S}\bar{y}} + a\bar{y} + \frac{1}{\delta+\lambda}(-aS + \lambda(c+d)) = \delta\Phi(1)\frac{1}{\delta}\left\{e^{\frac{\delta}{S}(1-\bar{y})} - \frac{\lambda}{\delta+\lambda}\right\}. \tag{43}$$

Using a Taylor expansion technique, under $\delta \to +0$, we get

$$\frac{1}{\delta}\left\{e^{-\frac{\delta}{S}(1-\bar{y})} - \frac{\lambda}{\delta+\lambda}\right\} = \frac{1}{\delta}\left\{1 + \frac{\delta}{S}(1-\bar{y}) - \left(1 - \frac{\delta}{\lambda}\right) + O(\delta^2)\right\} \to \frac{1}{S}(1-\bar{y}) + \frac{1}{\lambda}. \tag{44}$$

Therefore, by (44), we get the limit equation of (28) under $\delta \to +0$:

$$\frac{1-cS}{\lambda}e^{-\frac{\lambda}{S}\bar{y}} + c\left(1-\bar{y} + \frac{S}{\lambda}\right) + d = u\left\{\frac{1}{S}(1-\bar{y}) + \frac{1}{\lambda}\right\}. \tag{45}$$

Arranging (45) gives

$$(1-cS)e^{-\frac{\lambda}{S}\bar{y}} + c(\lambda(1-\bar{y}) + S) + d\lambda = u\left\{\frac{\lambda}{S}(1-\bar{y}) + 1\right\}. \tag{46}$$

Combining (42) and (46) yields

$$dS = (u - cS)(1-\bar{y}). \tag{47}$$

By (47), we should have

$$u > cS. \tag{48}$$

If (48) is true, then substituting (47) into (42) yields

$$(1-\bar{y})e^{-\frac{\lambda}{S}\bar{y}} = \frac{dS}{1-cS}. \tag{49}$$

The left-hand side of (49) is expressed as $F(\bar{y})$ using the strictly decreasing function $F:[0,1] \to \mathbb{R}$: $F(y) = (1-y)e^{-\frac{\lambda}{S}y}$. Since $F(0) = 1$ and $F(1) = 0$, we get the unique existence of $\bar{y} \in (0,1)$ if

$$0 < cS < 1 \quad \text{and} \quad 0 < \frac{dS}{1-cS} < 1, \tag{50}$$

the latter is rewritten as $(c+d)S < 1$ assuming the former. The former is trivially satisfied assuming the latter. The effective Hamiltonian $u$ is then found by substituting this $\bar{y}$ into (47). Notice that, by $\bar{y} \in (0,1)$, we can automatically verify (48) because of

$$u = cS + \frac{dS}{1-\bar{y}} > cS. \tag{51}$$

In summary, we get the following proposition.

## *Proposition 3*

*Assume $\bar{y} \in (0,1)$ for small $\delta$. We have the optimal control of the form (22) under the Ergodic limit $\delta \to +0$ if $(c+d)S < 1$.*

## *Remark 4*

The proof of the verification of the optimal control under the Ergodic limit is omitted because it essentially follows the proof of **Proposition 2** based on the boundedness (15) and uses the strategy similar to that in the prof of Theorem 3 of Wang (2001).

We also analyze parameter dependence of the threshold value $\bar{y}$ under the Ergodic limit. Because of the decreasing property of $F$, from (49), we can see that the optimal threshold $\bar{y}$ is decreasing with respect to $c, d$: namely, increasing the cost leads to a smaller threshold of sediment replenishment, which probabilistically leads to a less frequent sediment replenishment policy. The left-hand side of (49) is decreasing with respect to $\lambda$, leading to that $\bar{y}$ becomes smaller as well, which in this case is owing to the frequent (fine) observation collecting a larger amount of information.

Finally, the dependence on $S$ is analyzed. This case is a bit more complicated because the right-hand side of (49) is increasing with respect to $S$, and the left-hand side is also increasing with respect to $S$ for each fixed $\bar{y}$. Therefore, the above heuristic discussion does not apply in this case. We partially differentiate both-hand sides of (49) with respect to $S$ as

$$\left(-\frac{\partial \bar{y}}{\partial S}+(1-\bar{y})\left(-\frac{\partial \bar{y}}{\partial S}\frac{\lambda}{S}\right)+(1-\bar{y})\frac{\bar{y}\lambda}{S^2}\right)e^{-\frac{\lambda}{S}\bar{y}}=\frac{d}{(1-cS)^2}, \qquad (52)$$

which can be rewritten as

$$-\left(1+\frac{\lambda}{S}(1-\bar{y})\right)\frac{\partial \bar{y}}{\partial S}e^{-\frac{\lambda}{S}\bar{y}}+\frac{\bar{y}\lambda}{S^2}(1-\bar{y})e^{-\frac{\lambda}{S}\bar{y}}=\frac{d}{(1-cS)^2}. \qquad (53)$$

Substituting (49) into (53) yields

$$-\left(1+\frac{\lambda}{S}(1-\bar{y})\right)\frac{\partial \bar{y}}{\partial S}e^{-\frac{\lambda}{S}\bar{y}}+\frac{\lambda \bar{y}}{S}\frac{d}{1-cS}=\frac{d}{(1-cS)^2}. \qquad (54)$$

We then obtain

$$\left(1+\frac{\lambda}{S}(1-\bar{y})\right)\frac{\partial \bar{y}}{\partial S}e^{-\frac{\lambda}{S}\bar{y}}=\frac{d}{(1-cS)^2}\left(-1+\frac{\lambda \bar{y}}{S}(1-cS)\right)=\frac{d(-S+\lambda \bar{y}(1-cS))}{S(1-cS)^2} \qquad (55)$$

and thus

$$\frac{\partial \bar{y}}{\partial S}=C_0\left(\bar{y}-\frac{S}{\lambda(1-cS)}\right), \qquad (56)$$

where $C_0>0$. This means that for a small transport rate $S$ such that $\lambda>\frac{S}{\bar{y}(1-cS)}$, the threshold $\bar{y}$ should be increasing with respect to $S$, and vice versa. This means that a sufficiently high observation scheme assumes a stable sediment replenishment with a less depletion risk. The case $\lambda=\frac{S}{\bar{y}(1-cS)}$ is critical at which $\frac{\partial \bar{y}}{\partial S}=0$. However, it would not be so realistic from the standpoint of the original because $S$ is a function of the flow discharge and has been assumed to be stochastic. Consequently, the threshold level $\bar{y}$ is increasing (decreasing) for small (large) transport rate $S$. This unimodal nature of $\bar{y}$ on $S$ is validated numerically in Section 4.

### 3.1.3 Complete information case

A further model reduction is addressed in this sub-section. The last analysis of the exactly-solvable case focus on a full-information limit under the Ergodic case, which is the limit under $\delta \to +0$ and $\lambda \to +\infty$, the latter in particular means that the observation intensity is infinite (Pham and Tankov, 2008). The limit can be simply derived from the Ergodic limit, by taking the limit $\lambda \to +\infty$. Especially, we are interested in the existence of the non-trivial policy with

$\bar{y} \in (0,1)$.

Taking the limit $\lambda \to +\infty$ in (49) gives

$$\bar{y} = \frac{1-(c+d)S}{1-cS}. \tag{57}$$

By (57), the condition $\bar{y} \in (0,1)$ is satisfied if $0 < (c+d)S < 1$. Therefore, we see that the non-trivial optimal policy of the threshold type is still optimal if the sum of the fixed and proportional costs is sufficiently small. Otherwise, we encounter $\bar{y} \leq 0$ meaning that performing the most passive supplying policy $\bar{y} = 0$ or even doing nothing (no sediment supply) becomes optimal. Notice that the parameter dependence of the threshold $\bar{y}$ is qualitatively the same with that of the Ergodic case under the discrete observation.

## 3.2 Full problem

The full problem, which is the problem with generic $I \in \mathbb{N}$ is analyzed in this sub-section. If the value function is smooth ($\Phi_i \in C^1(D)$, $i \in M$), then a verification argument similar to that employed in the previous section applies and a threshold type control becomes optimal. Therefore, this case corresponds to a generalization of the exactly-solvable case.

Based on the mathematical analysis results of the exactly-solvable case, we assume the following regime-dependent threshold type control:

$$\eta^*_{\tau_k} = \begin{cases} 1 - Y_{\tau_k} & \left(0 \leq Y_{\tau_k} \leq \bar{Y}_{\alpha_{\tau_k}}\right) \\ 0 & \left(\bar{y}_{\alpha_{\tau_k}} < Y_{\tau_k} \leq 1\right) \end{cases}, \tag{58}$$

where $\bar{Y}_i \in (0,1)$, $i \in M$. Since the optimality equation in this case cannot be solved analytically like the exactly-solvable case analyzed above, the existence of the optimal policy is only an assumption. Therefore, we verify this optimal policy numerically in the next section, demonstrating that the threshold-type assumption is indeed reasonable.

The next proposition shows that the policy of the form (58), if it exists, is the optimal control and the associated smooth solution to the optimality equation is the value function.

*Proposition 4*

*Assume that there exists a function $\Psi_i \in C^1(D)$, $i \in M$ satisfying the optimality equation (16) pointwise, such that the associated candidate of an optimal control is (58). Then, this $\Psi$ is the value function $\Phi$ and the control (58) is optimal.*
**(Proof of Proposition 4)**

The proof is based on an application of Itô's formula, and is essentially the same with the **Proof of Proposition 2**. A difference is that now we must handle functions in $\Omega$. Therefore, we only present a sketch of the proof.

By Itô's formula, for $\psi_i \in C^1(D)$, $i \in M$ and $\bar{\eta} \in C$, we get

$$e^{-\delta T}\psi_{\alpha_T}(Y_T) - \psi_i(y) = \int_0^T e^{-\delta s}\left(-\delta\psi_{\alpha_{s-}}(Y_{s-}) - S\chi_{\{Y_{s-}>0\}}\psi'_{\alpha_{s-}}(Y_{s-}) - R_s\right)ds \\ + \sum_{0 \leq t \leq T} e^{-\delta t}\left(\psi_{\alpha_t}(Y_t + \eta_t dN_t) - \psi_{\alpha_t}(Y_t)\right), \quad T > 0 \quad (59)$$

with

$$R_s = \sum_{\alpha_{s-} \neq \alpha_s \in M} \nu_{\alpha_{s-},\alpha_s}\left(\Psi_{\alpha_{s-}}(Y_{s-}) - \Psi_{\alpha_s}(Y_{s-})\right). \quad (60)$$

By the assumption, $\Psi_i \in C^1(D)$, $i \in M$ and thus

$$e^{-\delta T}\psi_{\alpha_T}(Y_T) - \psi_i(y) \\ = \int_0^T e^{-\delta s}\lambda\left(\Psi_{\alpha_{s-}}(Y_s) - \inf_{\eta}\left\{\Psi_{\alpha_{s-}}(Y_{s-} + \eta) + c\eta + d\chi_{\{\eta>0\}}\right\}\right)ds + \int_0^T e^{-\delta s}\chi_{\{Y_{s-}=0\}}(-1)ds. \quad (61) \\ + \sum_{0 \leq t \leq T} e^{-\delta t}\left(\Psi_{\alpha_t}(Y_t + \bar{\eta}_t dN_t) - \Psi_{\alpha_t}(Y_t)\right)$$

The remaining part of the proof is essentially the same with that of **Proof of Proposition 2**.

### 3.3 Related issues

Several related issues like models with model uncertainty, models with delayed execution, and approaches from a viscosity viewpoint are discussed.

### 3.3.1 Model uncertainty

A comment on some advanced mathematical models considering model uncertainty is presented. The model uncertainty here means that not all the model parameters are accurately identified. For example, the transport rate is a physical quantity, but modern physical approaches still employ empirical laws (Ancey, 2020). The concept of nonlinear expectation (Neufeld and Nutz, 2017) harmonizes with the proposed stochastic control framework, and a problem with uncertain model parameter values can be formulated as a worst-case optimization problem having a saddle-point structure. This approach has successfully been applied to several stochastic control problems so far, especially in finance and economics (Neufeld and Nutz, 2018; Fouque and Ning, 2018).

Assume that the observation intensity $\lambda$ cannot be specified exactly, but only known to be in the compact $\Lambda = [\underline{\lambda}, \bar{\lambda}]$ with some constants $0 < \underline{\lambda} < \bar{\lambda} < \infty$. By the dynamic programming principle (Neufeld and Nutz, 2017), the optimality equation with this model ambiguity is formally derived as

$$\delta\Phi_i + S(q_i)\chi_{\{y>0\}}\frac{\mathrm{d}\Phi_i}{\mathrm{d}y} + \sum_{i\neq j\in M}\nu_{i,j}(\Phi_i - \Phi_j)$$
$$-\sup_{\lambda\in\Lambda}\left\{-\lambda\left(\Phi_i - \inf_{\eta\in C(y)}\{\Phi_i(y+\eta)+c\eta+d\chi_{\{\eta>0\}}\}\right)\right\}-\chi_{\{y=0\}}=0$$
, $(i,y)\in\Omega$, (62)

where the saddle-point nature appears in the non-local term.

In this case, we can fortunately simplify the worst-case optimality equation (62) by the relationship (20), from which we see that the quantity inside õsupö of (62) is not positive. Then, (62) reduces to the original optimality equation (16) with the reduced observation intensity $\lambda = \underline{\lambda}$. Therefore, the analysis presented in this paper applies in this case. Problems where the other model parameters are uncertain can be less trivial, and will be analyzed in our future works.

### 3.3.2 Delayed execution

Stochastic impulse control subject to delayed execution has been studied using the dynamic programming principle (Øksendal and Sulem, 2008; Perera and Long, 2017; Kharroubi et al., 2019; Bruder and Pham, 2009). The presented model can be extended to a delayed execution case in which there exists a time lag $\omega > 0$. For the sake of simplicity, assume that $\omega$ is a constant. The delayed execution may naturally arise when there exists a delay of decision-making in environmental management. In this case, (11) would be replaced by

$$Y_{(\tau_k+\omega)+} = Y_{\tau_k+\omega} + \eta_k \quad \text{with} \quad \eta_k = \begin{cases} 0 & (\text{Do nothing}) \\ 1-Y_{\tau_k} & (\text{Replenish}) \end{cases}. \quad (63)$$

Therefore, the decision-making result at $\tau_k$ affects the dynamics at the future time $\tau_k + \omega$. The original problem is recovered under $\omega \to +0$. The constraint $Y_t \in D$ ($t \geq 0$) is satisfied by (63).

According to the formulations of Perera and Long (2017) and Bruder and Pham (2009), the optimality equation in the delayed execution case formally becomes

$$\delta\Phi_i + S(q_i)\chi_{\{y>0\}}\frac{\mathrm{d}\Phi_i}{\mathrm{d}y}$$
$$+ \sum_{i\neq j\in M}\nu_{i,j}(\Phi_i - \Phi_j) + \lambda\left(\Phi_i - \inf_{\eta\in C(y)}L(\Phi_i,\eta)\right) - \chi_{\{y=0\}} = 0$$
, $(i,y)\in\Omega$ (64)

with

$$L(\Phi_i,\eta) = \mathbb{E}^{i,y}\left[\int_0^\omega e^{-\delta s}\chi_{\{Y_s=0\}}\,\mathrm{d}s + e^{-\delta\omega}\Phi_{\alpha_\omega}(Y_\omega + \eta)\right]. \quad (65)$$

The optimality equation involves another conditional expectation (65), which would have to be handled numerically by a Monte-Carlo method or a method based on the Feynman-Kac formula.

### 3.3.3 Viscosity solution approach

As discussed above, there exists a reasonable smooth solution to the optimality equation under certain assumptions. In general, solutions to degenerate elliptic equations are non-smooth and have non-differentiable points. Such solutions can be handled in the framework of viscosity solutions (Crandall et al., 1992). For problems with discrete and random observations, viscosity solutions can be defined following the previous research results (Pham and Tankov, 2008; Federico and Gassiat, 2014).

A difference between the conventional and present models is that the former handle smooth dynamics where the coefficients of the SDEs to be controlled are smooth, while it is not the case for the latter as shown in (4). The coefficients of the optimality equations (16) are discontinuous due the non-smooth system dynamics. For deterministic systems, this kind of degenerate elliptic equations can be analyzed from a viewpoint of viscosity solutions subject to discontinuous Hamiltonians (Barles et al., 2014). However, a difficulty may arise in our case because it is a stochastic case and is non-local. Fortunately, we could guess and verified the value function in the present model, and we therefore did not resort to employing the viscosity solution approach. Nevertheless, this approach can be useful when considering theoretical numerical analysis of the optimality equation in a viscosity sense (Barles and Souganidis, 1991).

## 4. Numerical computation
### 4.1 Numerical scheme

We employ the third-order Weight Essentially Non-Oscillatory (WENO) scheme based on the local Lax-Friedrichs finite difference discretization (Jiang and Peng, 2000). This is a high-resolution numerical scheme applied to a wide variety of problems. Its computational accuracy is third-order for solutions that are sufficiently smooth. Advantages of using this scheme are its simplicity and computational accuracy to handle nonlinear degenerate elliptic and hyperbolic problems such as the Hamilton-Jacobi type equations (Huang et al., 2008; Yoshioka et al., 2020b-c). A disadvantage is that the scheme is not necessarily monotone. In fact, from a mathematical viewpoint, it is better to employ a monotone, stable, and consistent scheme to guarantee convergence of numerical solutions in the viscosity sense (Barles and Souganidis, 1991). However, such schemes are usually at most first-order accurate, and are not always suited to applied problems. We do not use higher-order WENO reconstructions as well, because some of them do not get converged solutions when the true solutions are not smooth (Zhang et al., 2019).

The detail of the employed scheme is not presented here since it is found in Jiang and Peng (2000).

In what follows, we firstly check computational performance of the present scheme and then apply it to a realistic problem. We use the forward Euler pseud-temporal discretization method (Oberman, 2006) to obtain a steady numerical solution to the optimality equation. The initial guess is $\Phi \equiv 0$ and the integration period is $[0,T]$ with a sufficiently large $T > 0$.

### 4.2 Convergence property against exact solution

Computational accuracy of the numerical scheme is checked against the exact solution $\Psi$ in the single-regime case. The parameter values are set as follows: $S = 0.05$, $\delta = 0.2$, $c = 0.2$, $d = 0.3$, and $\lambda = 1/7$. The terminal time $T$ is set as 365/2, which is a sufficiently large value such that numerical solutions are close to time-independent at the terminal time. In fact, the absolute difference between numerical solutions in the successive time steps are smaller than $10^{-10}$ near the terminal time. The time step for the pseudo-temporal integration is chosen to be the sufficiently small value 1/800 for numerical stability. A bisection-like algorithm is applied to solving the nonlinear system (28), and the computed value 0.615195 is obtained up to the error smaller than $10^{-10}$. The corresponding exact solution is then constructed using (25). The domain $D$ is uniformly discretized with vertices as in the standard setting of finite difference schemes. The converged numerical solutions are obtained for the total number of vertices N = 51, 101, 201, 401, 801. The computed $\bar{y}$ with the finite difference scheme is assumed to be placed at a midpoint between successive vertices.

**Table 1** presents the computed errors measured by the standard $l^1$ error (mean of the errors between the exact and numerical solutions at all the vertices) and $l^\infty$ (maxim error between the exact and numerical solutions among all the vertices) error and the corresponding convergence rates. The computational results demonstrate that the numerical solutions successfully converge toward the exact solution and the convergence speed is second-order. The present scheme does not exhibit the expected third-order convergence possibly because of the lack of regularity of the exact solution $\Psi$: $\Psi \in C^1(D)$ but not always $\Psi \notin C^2(D)$ by (25).

**Table 2** presents the computed threshold value and the corresponding absolute error between the numerical and exact values. It seems that the threshold value is successfully approximated by the scheme. The obtained results demonstrate that the scheme can potentially discretize the optimality equation.

Although not presented here, using the local Lax-Friedrichs scheme with the same computational resolution can achieve only the first-order accuracy with the $l^\infty$ error larger than 5.50.E-03 and $l^1$ error larger than 3.40.E-03 when $N = 801$, which are several ten times larger errors than those with the WENO reconstruction. In addition, the scheme with the WENO reconstruction achieves higher accuracy when $N = 201$. The computational results imply usefulness of the scheme. It should be noted that the scheme is convergent irrespective to the use the WENO reconstruction.

**Table 1.** Computed errors measured by the standard $l^1$ and $l^\infty$ errors and the corresponding convergence rates. The convergence rate between the errors $e_1$ and $e_2$ of the resolutions $N = N_1$ and $N = N_2$ ($N_1 < N_2$) is calculated as $\log_{N_2/N_1}(e_1/e_2)$.

| N | 51 | 101 | 201 | 401 | 801 |
|---|---|---|---|---|---|
| $l^\infty$ error | 1.98.E-02 | 5.58.E-03 | 1.52.E-03 | 4.00.E-04 | 1.10.E-04 |
| $l^1$ error | 5.59.E-03 | 1.45.E-03 | 3.80.E-04 | 9.59.E-05 | 2.40.E-05 |
| $l^\infty$ convergence rate | 1.9 | 1.9 | 1.9 | 1.9 | |
| $l^1$ convergence rate | 2.0 | 1.9 | 2.0 | 2.0 | |

**Table 2.** Computed threshold value and the corresponding absolute error between the numerical and exact values.

| N | 51 | 101 | 201 | 401 | 801 |
|---|---|---|---|---|---|
| Computed $\bar{y}$ | 0.61 | 0.615 | 0.6175 | 0.61625 | 0.615625 |
| Error | 5.20.E-03 | 1.95.E-04 | 2.31.E-03 | 1.05.E-03 | 4.30.E-04 |

### 4.3 Realistic case

A more realistic case is considered where the model parameters are identified from the available record and a hydraulic formula. The focus here is a model application to a downstream environment of an existing river in Japan (O Dam, H River, Japan). The O Dam has been working from 2011. Since then, the transported sediment from the upstream was trapped by the dam.

In this river, a local fishery cooperative and the Ministry of Land, Infrastructure, Transport and Tourism and are playing the role of environmental manager of the dam-downstream river environment. They, for the first time in this river, experimentally replenished the sediment in April 2020 with the amount of 100 (m$^3$). However, the observation intensity and the sediment replenishment scheme have not been determined so far. Our application thus concerns an emerging case of the control problem.

In what follows, the presented model with identified model parameter values are applied to this realistic problem. The parameters on the system dynamics are identified from the available data and semi-empirical physical laws. On the other hand, those on the performance index and observation process should depend on the decision-maker. Here, we determine these parameter values considering the time-scale of the system dynamics and decision-making.

### 4.3.1 Markov chain

The model application area is the just downstream reach of O dam, where the river flow discharge in this area can be identified as the outflow discharge from the dam, as in the previous research (Yoshioka et al. (2020a, 2020c)). The matrix and the total number of regimes of the flow Markov chain have been identified using a maximum entropy principle based on a public hourly dam operation data (outflow discharge) from April 2016 to March 2020: $I = 42$ with the discharge for each flow regime $q_i = 1.25 + 2.5i$ ($i = 0, 1, 2, ..., 42$).

**Figure 1** plots the identified matrix $\nu$, which is utilized in what follows. Historically, the maximum outflow discharge of the dam exceeds 300 (m³/s) several times in each year, but such events are of less importance in the computation below because the average outflow discharge during this period was estimated as 5.01 (m³/s) with the standard deviation 15.4 (m³/s). In fact, occurrence probability of an event exceeding the outflow discharge of $q_I$ is less than 0.4 % according to the estimated Markov chain. See, **Figure 2** for the stationary probability density $p = \{p_i\}_{i \in M}$. The condition (3) is satisfied because we numerically have $p_i > 0$ for all $i \in M$ and the all the regimes are transient.

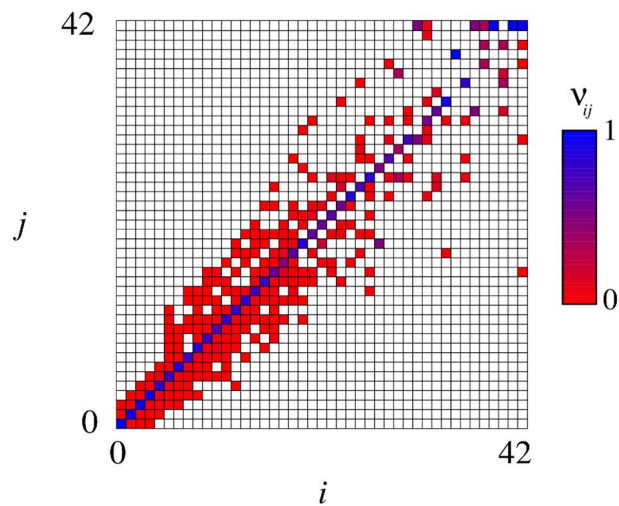

**Figure 1.** Identified matrix $\nu$ using the hourly outflow discharge.

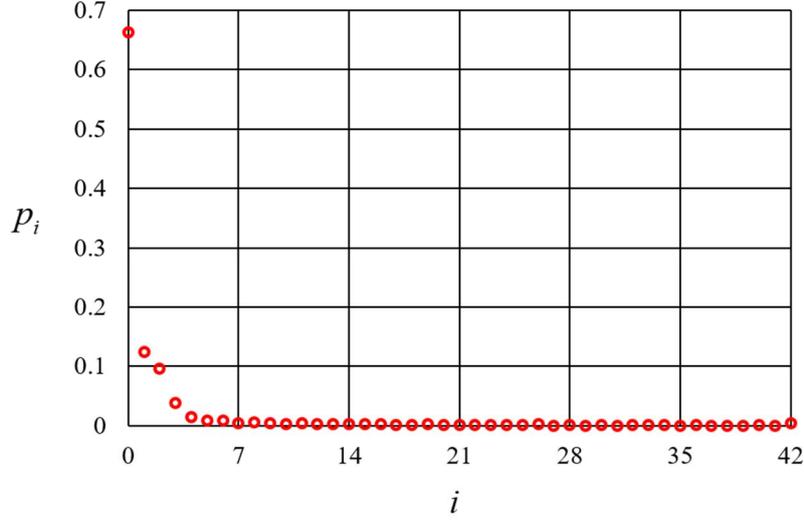

**Figure 2.** Computed stationary probability density $p = \{p_i\}_{i \in M}$ of the identified Markov chain.

### 4.3.2 Transport rate

The transport rate, which is the amount of transported sediment per unit time, is estimated using the widely-used Meyer-Peter-Müller formula (Meyer-Peter and Müller, 1948; Wong and Parker, 2006). The formula in our context is expressed as

$$S(q) = 8B\gamma^{1.5}\sqrt{g\sigma}\max\{\theta - \theta_c, 0\}^{\frac{3}{2}}, \quad \theta = \frac{\tau(q)}{\rho\sigma g\gamma}, \text{ and } \theta_c = 0.047, \quad (66)$$

where $B$ is the river channel width, $g$ is the gravitational acceleration, $\gamma$ is the diameter of sediment particles, $\rho$ is the density of water, $\sigma = \frac{\rho_s}{\rho} - 1 > 0$ with $\rho_s$ the density of soil particles, $\tau$ is the bed shear stress as a positive bounded, and increasing function of the discharge $q > 0$. The formula (66) suggests that, given some sediment with prescribed physical quantities, the sediment transport will occur only when the discharge is sufficiently large such that $\theta > \theta_c$. To complete the formula (66), the Manning's formula of the bottom shear stress under a uniform flow assumption is utilized (Chapter 1 of Szymkiewicz, 2010):

$$\tau(q) = \rho g h l = \rho g n^{\frac{3}{5}} l^{\frac{7}{10}} B^{-\frac{3}{5}} q^{\frac{3}{5}}, \quad (67)$$

where $n = O(10^{-2})$ is the Manning's roughness coefficient and $l$ is the channel slope. We can predict the transport rate at each regime by using the formulae (66)-(67) with a set of prescribed values of the physical quantities. Therefore, the transport rate is scaled with the discharge as $S(q) \propto q^{\frac{9}{10}}$ for large $q$, implying its almost linear dependence on $q$.

Recall that the sediment storage has been normalized in $D = [0,1]$, which is considered as a non-dimensional zed physical domain of the storage $\bar{D} = [0, \bar{Y}]$, where $\bar{Y}$ is the total volume of sediment (m³) storable in the reach. In this case, the form of the SDE(4) is unchanged, and the range of the variable $Y$ becomes $\bar{D}$.

In the numerical computation, we use the following parameter values covering typical river environmental condition and sediment material properties: $g = 9.81$ (m/s²), $B = 25$ (m), $l = 0.001$ (m), $n = 0.035$ (m$^{1/3}$/s), $\rho = 1,000$ (kg/m³), $\rho_s = 2,600$ (kg/m³), $\gamma = 5.0 \times 10^{-3}$ (m), and $\bar{Y} = 100$ (m³). Under this parameter setting, we have $S_i = 0$ for $i = 0, 1$, meaning that the sediment transport toward does not occur during these low flow regimes.

### 4.3.3 Other parameters

Parameter values involved in the performance index $\phi$ of (13) have to be specified for the numerical computation. There are the three parameters $c, d, \delta$. The discount rate $\delta$ is set as 0.2 (1/day) assuming that the time-scale for the decision-making, which is $\delta^{-1}$ is $O(1)$ (day). This means that the decision-making is assumed to has a daily time-scale. Values of the other parameters are specified as $c = 0.02$ and $d = 0.01$. These parameter values are determined by a trial and error approach so that a non-trivial (non-constant) optimal policy, which is of our interest, is obtained. In fact, specifying a too large (resp., too small) $c$ or $d$ leads to the policy that does not supply sediment at all (always supply the sediment) at each observation. Finally, the observation intensity is set as and $\lambda = 1/7$ (1/day) assuming an observation process having one observation in each week on average.

### 4.3.4 Computational resolution

The terminal time $T$ is set as 90 (day), which is a sufficiently large value such that numerical solutions are sufficiently close to time-independent at the terminal time. The time step for the temporal integration is chosen as 0.000025 (day). The domain $D$ is discretized uniformly with $N = 301$ vertices. Choosing this sufficiently small time step is due to the large sediment transport rate $S$ for the regimes close to $i = I$. In fact, $S = O(10^2)$ (1/day) in such regimes. This fact combined with $N = O(10^2)$ leads to the maximally allowable time step as $O(10^{-4})$. The absolute difference between numerical solutions in the successive time steps is smaller than $10^{-9}$ near the terminal time.

### 4.3.5 Computational results

We numerically analyze computed optimal controls focusing on its dependence on the discount rate determined by the decision-maker and the flow regimes.

**Figure 3** shows the computed value function $\Phi = \Phi(i, y)$ and the associated optimal control $\eta = \eta^*$ in the domain $\Omega$. The computational results suggest that the numerical solutions are successfully obtained without visible spurious oscillations. The computed optimal control is a threshold type (22), numerically validating the assumption made in the mathematical analysis in the previous section. The decision-maker can decide whether he/she should carry out the replenishment at each observation time based on this threshold type control.

The above-presented computational results suggest that analyzing the optimal control problem reduces to investigating behavior of the threshold level $\overline{Y} = \{\overline{Y}_i\}_{i \in M}$: the free boundary. An important point to be considered from an engineering viewpoint is dependence of the optimal policy on the observation intensity: namely, dependence of $\overline{Y}$ on $\lambda$. We therefore numerically solve the optimality equation (16) for different values of $\lambda$ and compare the free boundaries $\overline{Y}$ among the different cases.

**Figure 4** shows the computed $\overline{Y}$ for different values of $\lambda$. The free boundary $\overline{Y} = \{\overline{Y}_i\}_{i \in M}$ is monotonically decreasing with respect to the observation intensity $\lambda$. This means that the management policy with a less intensive observation process should set a larger threshold value, so that he/she encounter the sediment depletion less frequently. The computational results also give an important implication at relatively low flow regimes where $i$ is small. The difference among the free boundaries for different values of $\lambda$ is less significant for the not small $i \geq 6$; $\overline{Y}_i$ in these regimes increase at most 0.05 as $\lambda$ decreases from 1/1 to 1/30. Since $\lambda$ represents the inverse time scale of the observation interval, this implies that the observation frequency is of less importance if the sediment replenishment is carried out at the relatively high flow regimes. However, as shown in **Figure 2**, the occurrence of such regimes is significantly less than that of the relatively low flow regimes with $i \leq 5$. Therefore, a suggestion obtained from this sensitivity analysis is that the threshold of the sediment replenishment, the free boundary, should be carefully designed especially for the low flow regimes if the decision-maker is considering the policies with different observation intensities.

The transport rate $S$ is now regime-dependent, but **Figure 4** suggests that the free boundary $\overline{Y} = \{\overline{Y}_i\}_{i \in M}$ as a function of the regime $i$ is increasing for small $i$ (small $S$), while it is decreasing for large $i$ (large $S$). This profile of $\bar{Y}$ suggests that the unimodal dependence on $S$ in the exactly-solvable case is inherited in this regime-switching case, implying the usefulness of the simplified model.

We also consider behavior of the free boundary $\bar{Y}$ for different values of $\lambda$ under the Ergodic case ($\delta \to +0$): the long-run limit. The Ergodic case is computationally handled by setting $\delta = 0$ and a sufficiently large terminal time $T$ such that the computed $\eta^*$ is close to be time-independent at the terminal time (the large $T$-method (Qian, 2003)). We preliminary checked that choosing $T = 90$ (day) is sufficiently large for computing the Ergodic limit.

**Figure 5** shows the computed value function $\Phi$ and the associated optimal control $\eta^*$ for the Ergodic case with $\lambda = 1/7$ (1/day). The computed $\Phi$ is almost constant for each regime and that the threshold type control is still optimal. The analysis results are consistent with the exactly-solvable case in **Section 3**, suggesting that the exactly-solvable case can capture the essential property of the optimal policy despite its simplicity. Especially, again the free boundary $\bar{Y} = \{\bar{Y}_i\}_{i \in M}$ has the unimodal nature for the regime $i$.

Finally, **Figure 6** shows the computed $\bar{Y}$ for different values of $\lambda$ under the Ergodic case. Sensitivity of the free boundary $\bar{Y}$ on the observation intensity $\lambda$ is smaller than the discounted case presented above. The computational results suggest that the decision-maker considering the sediment storage management in the long-run should follow the threshold type control but with less care on the threshold values of the relatively low flow regimes.

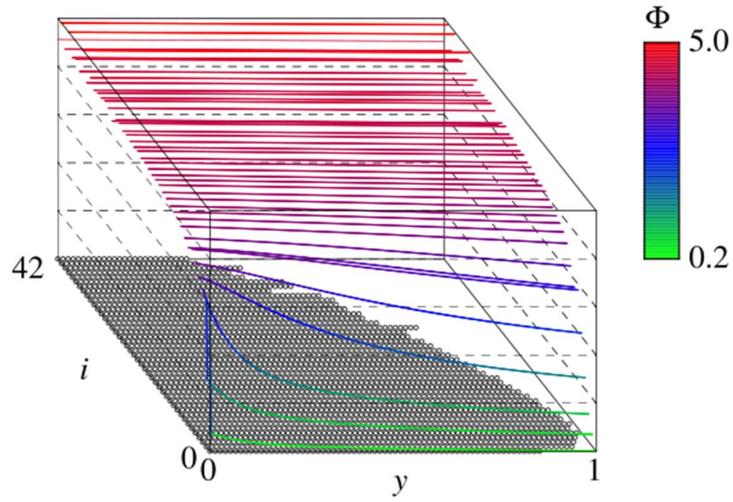

**Figure 3.** The computed value function $\Phi$ and the associated optimal control $\eta^*$.

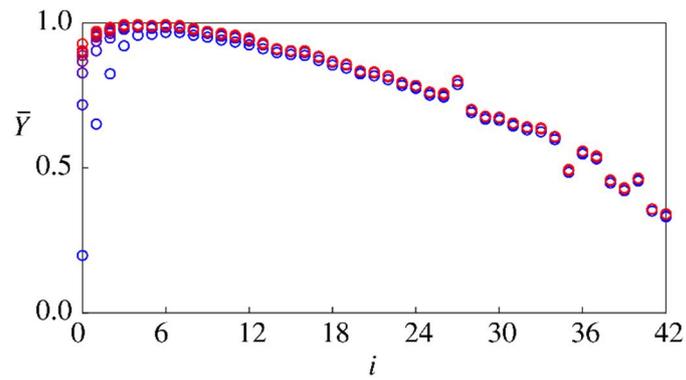

**Figure 4.** The computed free boundaries $\bar{Y} = \{Y_i\}_{i \in M}$ for different values of the observation intensity $\lambda$ ($\lambda = 1/1, 1/3, 1/5, 1.7, 1/9, 1/11, 1/13$, and $1/30$: the colors are from Blue to Red in this order). The free boundary $\bar{Y}$ moves upward in the figure panel as $\lambda$ decreases.

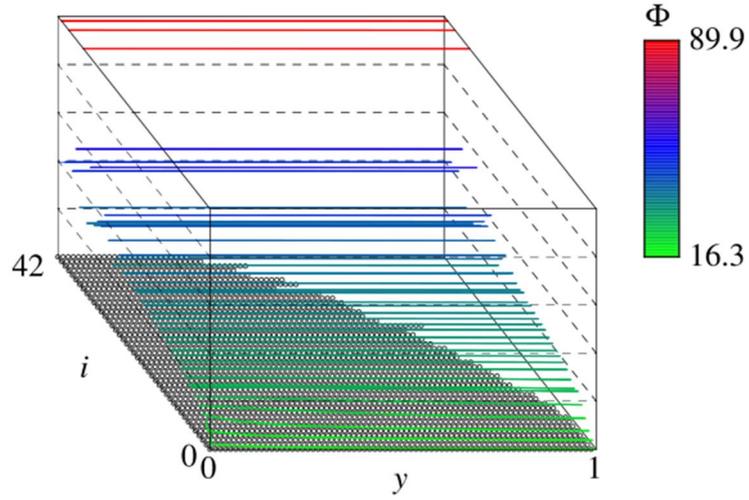

**Figure 5.** The computed value function $\Phi$ and the associated optimal control $\eta^*$ (Ergodic case where $\lambda = 1/7$ (1/day) and $\delta = 0$ (1/day))

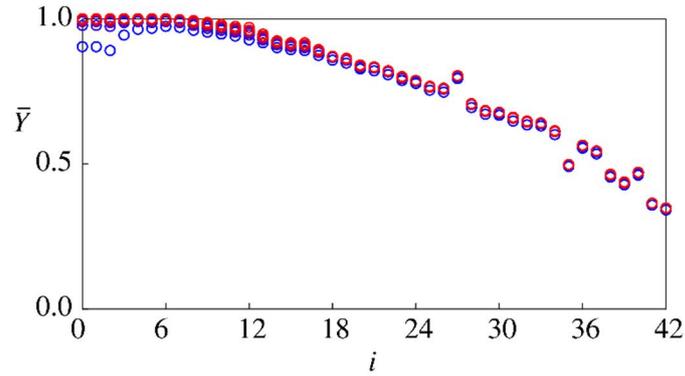

**Figure 6.** The computed free boundaries $\overline{Y} = \{Y_i\}_{i \in M}$ for different values of the observation intensity $\lambda$ (Ergodic case where $\delta = 0$ (1/day). We examine $\lambda = 1/1, 1/3, 1/5, 1.7, 1/9, 1/11, 1/13,$ and $1/30$: the colors are from Blue to Red in this order). The free boundary $\overline{Y}$ moves upward in the figure panel as $\lambda$ decreases as in **Figure 4**.

## 5. Conclusions

We formulated a cost-efficient sediment replenishment problem in a stochastic river environment as an optimal control problem with discrete and random observation/intervention. The system dynamics and the performance index are non-smooth and the optimality equation to find the most cost-effective management policy inherits this property in the coefficients. We could exactly solve the optimality equation under the single-regime case, and derived a similar verification result on the full problem. For the exactly-solvable case, we obtained closed-form solutions under simplified cases including Ergodic and complete information cases.

We also numerically analyzed the optimality equation and the optimal policy using a finite difference scheme equipped with the WENO reconstruction. Convergence property of the scheme was checked against the exact solution. The parameter values and coefficients in the model are identified using the available data and hydraulic formulae. The computational results suggest that the threshold type policy is indeed optimal under the realistic case.

We could fortunately find a smooth solution to the optimality equation, but solutions would not be sufficiently smooth in cases that can be more complicated. A possible option to deal with such cases is to use a variational framework that has been successfully applied to the degenerate elliptic problem associated with non-smooth dynamics (Bensoussan et al., 2016). Flood disturbance can also be handled using some SDEs (Biao et al., 2016; Ferrazzi and Botter, 2019) with which climate change effects on the river flow regimes can be parameterized more in detail than the Markov chain approach. Employing a state aggregation technique (Parpas and Webster, 2014) may lead to a simpler model without critically degrading the essential dynamics. In this paper, the complete information case was theoretically analyzed for the exactly-solvable case, while it was not in the numerical computation. A difficulty was the explicit nature of the employed scheme where the time increment for the pseudo-temporal integration is required to be taken extremely small if we specify a large observation intensity. An implicit numerical scheme should be employed for resolving this issue. The related issues discussed in Section 3 are also worth investigating because they are closely related to both theoretical analysis and applications.

**Acknowledgements**

JSPS Research Grant No. 19H03073, Kurita Water and Environment Foundation Grant No. 19B018, a grant for ecological survey of a life history of the *landlocked ayu Plecoglossus altivelis altivelis* from the Ministry of Land, Infrastructure, Transport and Tourism of Japan, and a research grant for young researchers in Shimane University support this research.